\documentstyle[prd,aps,preprint,tighten,epsfig]{revtex}

\begin{document}

\draft

\title{Sum rules of four-neutrino mixing in matter}

\author{\bf He Zhang}
\address{CCAST (World Laboratory), P.O. Box 8730, Beijing 100080, China \\
and Institute of High Energy Physics, Chinese Academy of Sciences, \\
P.O. Box 918 (4), Beijing 100049, China
\footnote{Mailing address} \\
({\it Electronic address: zhanghe@mail.ihep.ac.cn}) } \maketitle

\begin{abstract}

Assuming the existence of one light sterile neutrino, we
investigate the neutrino flavor mixing matrix in matter. Sum rules
between the mixing parameters in vacuum and their counterparts in
matter are derived. By using these new sum rules, we obtain the
simple but exact expressions of the effective flavor mixing matrix
in matter in terms of neutrino masses and the mixing parameters in
vacuum. The rephasing invariants, sides of unitarity quadrangles
and oscillation probabilities in matter are also achieved. Our
model-independent results will be very helpful for analyzing
flavor mixing and CP violation in the future long-baseline
neutrino oscillation experiments.

\end{abstract}

\newpage

\section{Introduction}

The facts that neutrinos are massive and lepton flavors are mixed
have been confirmed by recent solar \cite{SNO}, atmospheric
\cite{SK}, reactor \cite{KM} and accelerator \cite{K2K} neutrino
oscillation experiments. However, the data coming from the Liquid
Scintillation Neutrino Detector (LSND) experiment\cite{LSND} have
neither been confirmed nor refuted by the MiniBooNe
experiment\cite{MiniBooNE}. If the LSND result is proved to be
true, most probably there exist more than three neutrino mass
eigenstates. From the phenomenological point of view, the most
straightforward and attractive way to understand the result of
LSND is to introduce the sterile neutrinos, which are electroweak
singlets and do not couple to the Standard Model $W$ and $Z$
bosons. Such sterile neutrinos mix with the active neutrinos, thus
they can significantly influence neutrino oscillations. The
sterile neutrinos may also be the candidates of warm dark
matter\cite{dark} in cosmology. Due to the Mikheyev, Smirnov and
Wolfenstein (MSW) effect\cite{MSW} within the supernovae, visible
effects of sterile neutrinos in supernova explosion will probably
be detected in the forthcoming neutrino telescopes\cite{super}.
There are also plenty of other reasons to believe the possible
existence of sterile neutrinos\cite{white}. Even if the LSND
experiment is not confirmed, sterile neutrinos can not be fully
excluded, and the number of neutrino species is still an important
open question in neutrino physics.

The direct effect of introducing sterile neutrinos is the mixing
between active neutrinos and sterile neutrinos. For simplicity,
here we only consider one light sterile neutrino. Hence the
neutrino mass eigenstates will become linear combinations of three
active neutrinos ($\nu^{}_e,\nu^{}_\mu,\nu^{}_\tau$) and one
sterile neutrino ($\nu^{}_s$). To be explicit, we can write
\begin{eqnarray}
\left ( \matrix{ \nu^{}_s \cr \nu^{}_e \cr \nu^{}_\mu \cr
\nu^{}_\tau \cr} \right ) \; = \; \left ( \matrix{ V^{}_{s0} &
V^{}_{s1} & V^{}_{s2} & V^{}_{s3} \cr V^{}_{e0} & V^{}_{e1} &
V^{}_{e2} & V^{}_{e3} \cr V^{}_{\mu 0} & V^{}_{\mu 1} & V^{}_{\mu
2} & V^{}_{\mu 3} \cr V^{}_{\tau 0} & V^{}_{\tau 1} & V^{}_{\tau
2} & V^{}_{\tau 3} \cr} \right ) \left ( \matrix{ \nu^{}_0 \cr
\nu^{}_1 \cr \nu^{}_2 \cr \nu^{}_3 \cr} \right ) \; ,
\end{eqnarray}
where $\nu^{}_i$ ($i=0,1,2,3$) denote the neutrino mass
eigenstates, and $V^{}_{\alpha i}$ ($\alpha=s,e,\mu,\tau$) is the
lepton flavor mixing matrix.

The behavior of neutrino propagation in vacuum can be described by
the effective Hamiltonian
\begin{eqnarray}
{\cal H}  =  \frac{1}{2E} \left (V D^2 V^{\dagger} \right ) \ ,
\end{eqnarray}
where $D \equiv {\rm Diag}\{m^{}_0, m^{}_1, m^{}_2, m^{}_3 \}$,
$m^{}_i$ are the masses of $\nu^{}_i$ and $E$ denotes the neutrino
beam energy. In order to measure the lepton flavor mixing angles
precisely and to answer the question whether neutrinos violate CP,
future accelerator experiments with very long baselines\cite{Long}
are needed. In such long baseline experiments, neutrino beams
should travel through the terrestrial matter, and their
propagation can be significantly modified when they interact with
matter. The effective Hamiltonian responsible for neutrinos
travelling in matter can be written in a form analogous to Eq. (2)
\begin{eqnarray}
 \tilde{\cal H} & = &  \frac{1}{2E} \left (\tilde{V} \tilde{D}^2
\tilde{V}^{\dagger} \right ) \ ,
\end{eqnarray}
where $\tilde{D} \equiv {\rm Diag}\{\tilde{m}^{}_0,
\tilde{m}^{}_1, \tilde{m}^{}_2, \tilde{m}^{}_3 \}$,
$\tilde{m}^{}_i$ are the effective neutrino masses in matter and
$\tilde{V}$ denotes the effective lepton mixing matrix in matter.
$\tilde{\cal H}$ is related with $\cal H$ as
\begin{eqnarray}
\tilde{\cal H}-{\cal H} = \frac{1}{2E}{\cal A} \ ,
\end{eqnarray}
where ${\cal A}=2E \times {\rm Diag} ( a',a,0,0 )$ with the matter
parameters $a'=\sqrt{2}G^{}_{\rm F} N^{}_n/2$ and
$a=\sqrt{2}G^{}_{\rm F} N^{}_e$ arising from coherent forward
neutrino scattering\cite{Grimus}. Here $G^{}_{\rm F}$ is the Fermi
constant, $N^{}_e$ and $N^{}_n$ are the numbers of electrons and
neutrons per unit volume. For most realistic long-baseline
neutrino experiments, the background density can usually be
treated as a constant\cite{M} ($N^{}_e=\rm constant$ and
$N^{}_n=\rm constant$). In this work, we take $N^{}_{e}$ and
$N^{}_{n}$ as constants.

In respect that the phenomenology of neutrino oscillations in
matter can be formulated in terms of $\tilde{V}$ and
$\tilde{m}^{}_i$, a number of authors have discussed the
analytical relations between $\tilde{V}$ and
$V$\cite{Barger,Zaglauer,Xing00-01,Freund,Kimura,Smirnov,Zhang05,others}
in the framework of three active neutrinos. The main purpose of
this paper is to derive the exact analytical relationships between
$\tilde{V}$ and $V$ in the four-neutrino mixing scheme. After an
extension of our previous results\cite{Zhang05} from three active
neutrinos to three active neutrinos and one sterile neutrino, the
model-independent sum rules between ($\tilde{V}, \tilde{m}^{}_i$)
and ($V, m^{}_i$) are given in section II. The relations between
moduli $|\tilde{V}^{}_{\alpha i}|$ and $|V^{}_{\alpha i}|$ are
then obtained in section III, and so are the relations between
$\tilde{V}^{}_{\alpha i}\tilde{V}^{\ast}_{\beta i}$ and
$V^{}_{\alpha i}V^{\ast}_{\beta i}$. Applications of our relations
are also shown in section III. Finally, a brief summary is
presented in section IV.

\section{sum rules in the four-neutrino mixing scheme}

The effective Hamiltonians $\cal H$ and $\tilde{\cal H}$ in Eqs.
(2) and (3) can be expressed as
\begin{eqnarray}
{\cal H}^{}_{\alpha \beta} & = &  \frac{1}{2E} \sum_{i} m^2_i
V^{}_{\alpha i} V^{\ast}_{\beta i} \ ,
\nonumber \\
\tilde{\cal H}^{}_{\alpha \beta} & = &  \frac{1}{2E} \sum_{i}
\tilde{m}^2_i \tilde{V}^{}_{\alpha i} \tilde{V}^{\ast}_{\beta i} \
,
\end{eqnarray}
where the Greek indices $\alpha$, $\beta$ and so on generally run
over $s,e,\mu$ and $\tau$, and the Latin indices $i$, $j$, etc.
run over $0$, $1$, $2$ and $3$ here after. Eq. (4) turns out to be
\begin{eqnarray}
\sum_{i} \tilde{m}^2_i \tilde{V}^{}_{\alpha i}
\tilde{V}^{\ast}_{\beta i}  & = & \sum_{i} m^2_i V^{}_{\alpha i}
V^{\ast}_{\beta i} + {\cal A}^{}_{\alpha \beta}
\end{eqnarray}
with
\begin{eqnarray}
{\cal A}^{}_{\alpha \beta} = a' \delta^{}_{s \alpha} \delta^{}_{s
\beta} + a \delta^{}_{e \alpha} \delta^{}_{e \beta} \ .
\end{eqnarray}
Here only the ($s,s$) and ($e,e$) elements of $\cal A$ are not
vanishing. The unitarity conditions of $V$ and $\tilde{V}$ yield
\begin{eqnarray}
\sum_{i}  V^{}_{\alpha i} V^{\ast}_{\beta i}  = \sum_{i}
\tilde{V}^{}_{\alpha i} \tilde{V}^{\ast}_{\beta i} =
\delta^{}_{\alpha\beta} \ .
\end{eqnarray}
These two sets of sum rules are straightforward and obvious, but
they are not enough for our calculations. In order to obtain the
relations between $\tilde{V}^{}_{\alpha i} \tilde{V}^{\ast}_{\beta
i}$ and $V^{}_{\alpha i} V^{\ast}_{\beta i}$, we need more linear
independent equations.

Taking into account the unitarity of $V(\tilde{V})$, we find that
$({\cal H})^n=(1/2E)^n V D^{2n} V^{\dagger} $ and $(\tilde{\cal
H})^n=(1/2E)^n \tilde{V} \tilde{D}^{2n} \tilde{V}^{\dagger} $
hold, in which $n=0,\pm1,\pm2$, etc. To be explicit, we obtain the
elements of ${\cal H}(\tilde{\cal H})$ as
\begin{eqnarray}
({\cal H}^{n})_{\alpha \beta} & = & \frac{1}{(2E)^n} \sum_{i}
m^{2n}_{i} V^{}_{\alpha
i} V^{\ast}_{\beta i} \ , \nonumber \\
(\tilde{\cal H}^{n})^{}_{\alpha \beta} & = &
\frac{1}{(2E)^n}\sum_{i} \tilde{m}^{2n}_{i} \tilde{V}^{}_{\alpha
i} \tilde{V}^{\ast}_{\beta i} \ .
\end{eqnarray}
Note that the case $n=0$ corresponds to the unitarity conditions,
and $n=1$ corresponds to the sum rules in Eq. (6). For the case
$n=2$, we square the two sides of Eq. (6) and obtain
\begin{eqnarray}
\sum_{i} \tilde{m}^{4}_{i} \tilde{V}^{}_{\alpha i}
\tilde{V}^{\ast}_{\beta i} = \sum_{i} \left[ m^{4}_{i} + m^{2}_{i}
\left( {\cal A}^{}_{\alpha \alpha} +{\cal A}^{}_{\beta \beta}
\right) \right] V^{}_{\alpha i} V^{\ast}_{\beta i} + {\cal
A}^{2}_{\alpha \beta} \ .
\end{eqnarray}
In the calculation of Eq. (10), the expression of $\cal A$ in Eq.
(7) has been considered. For the case $n=3$, we get the forth set
of equations by cubing both two sides of Eq. (6)
\begin{eqnarray}
\sum_{i} \tilde{m}^{6}_{i} \tilde{V}^{}_{\alpha i}
\tilde{V}^{\ast}_{\beta i} & =&  \sum_{i} \left[ m^{6}_{i} +
m^4_{i} \left( {\cal A}^{}_{\alpha \alpha} + {\cal A}^{}_{\beta
\beta} \right)+m^{2}_{i} \left( {\cal A}^{2}_{\alpha \alpha}+
{\cal A}^{2}_{\beta \beta} + {\cal A}^{}_{\alpha \alpha} {\cal
A}^{}_{\beta \beta} \right)\right] V^{}_{\alpha i} V^{\ast}_{\beta
i}
\nonumber \\
&&+ \sum_{\gamma,i,j} m^{4}_{i} V^{}_{\alpha i}  V^{}_{\gamma j}
V^{\ast}_{\gamma i} V^{\ast}_{\beta j} {\cal A}^{}_{\gamma \gamma}
+ {\cal A}^{3}_{\alpha \beta}  \ .
\end{eqnarray}

Eqs. (6), (8), (10) and (11) constitute a full set of linear
equations of $\tilde{V}^{}_{\alpha i} \tilde{V}^{\ast}_{\beta i}$.
By solving these equations, one can derive the exact expressions
of $\tilde{V}^{}_{\alpha i} \tilde{V}^{\ast}_{\beta i}$ in terms
of $\tilde{m^{}_{i}}$, $m^{}_i$ and $V^{}_{\alpha i}
V^{\ast}_{\beta i}$. Note that, the sum rules obtained above are
only valid for neutrinos. As for antineutrinos, the corresponding
sum rules can directly be obtained by replacing ($V$, $\tilde{V}$,
$\cal A$) with ($V^{\ast}$, $\tilde{V}^{\ast}$, $-\cal A$). Our
formulae can be generalized to include more than one sterile
neutrinos. In this paper, we only concentrate on the four-neutrino
mixing scheme and the main results will be given in the next
section.

\section{the mixing matrix elements in matter}

The sum rules obtained in the previous section totally contain ten
sets of linear equations, which are related to the ten independent
elements in $\tilde{ \cal H}$. Let us first consider the case
$\alpha=\beta$, in which the sum rules are simplified to
\begin{eqnarray}
\sum_{i} |\tilde{V}^{}_{\alpha
i}|^2 & = & \sum_{i}  |V^{}_{\alpha i}|^2 = 1 \ , \nonumber \\
\sum_ {i} \tilde{m}^2_i |\tilde{V}^{}_{\alpha i}|^2  & = &
\sum_{i} m^2_i |V^{}_{\alpha i}|^2 + {\cal A}^{}_{\alpha \alpha} \
,
\nonumber \\
\sum_{i} \tilde{m}^4_i |\tilde{V}^{}_{\alpha i}|^2  & = & \sum_{i}
\left[ m^{4}_{i} + 2 m^{2}_{i}  {\cal A}^{}_{\alpha \alpha}
\right] |V^{}_{\alpha i}|^2  + {\cal A}^{2}_{\alpha \alpha} \ ,
\nonumber \\
\sum_{i} \tilde{m}^{6}_{i} |\tilde{V}^{}_{\alpha i}|^2 & =&
\sum_{i} \left[ m^{6}_{i} + 2 m^4_{i} {\cal A}^{}_{\alpha \alpha}
+ 3 m^{2}_{i} {\cal A}^{2}_{\alpha \alpha}\right] |V^{}_{\alpha
i}|^2  \nonumber
\\
&& + \sum_{\gamma,i,j} m^{4}_{i} V^{}_{\alpha i} V^{}_{\gamma j}
V^{\ast}_{\gamma i} V^{\ast}_{\alpha j} {\cal A}^{}_{\gamma
\gamma} + {\cal A}^{3}_{\alpha \alpha}  \ .
\end{eqnarray}
After a lengthy calculation, the solutions of Eq. (12) are
obtained as
\begin{eqnarray}
|\tilde{V}^{}_{\alpha i}|^2  = \frac{1}{\prod \limits_{k \neq i}
\tilde{\Delta}_{ik}} \sum_{j} \left( F^{i j}_{\alpha }+
C^{}_{\alpha} \right) |V^{}_{\alpha j}|^2 \ ,
\end{eqnarray}
where
\begin{eqnarray}
F^{ij}_{\alpha} & = & \prod \limits_{k \ne i} \left( {\cal
A}^{}_{\alpha \alpha} + \widehat{\Delta}^{}_{j k}\right) \ ,
\nonumber \\
C^{}_{\alpha} & = & -\frac{1}{2} \sum_{\gamma,m,n} \Delta^2_{mn}
V^{}_{\alpha m}  V^{}_{\gamma n}  V^{\ast}_{\gamma m}
V^{\ast}_{\alpha n} {\cal A}^{}_{\gamma \gamma} \ .
\end{eqnarray}
Here we have defined the mass-squared differences $\Delta^{}_{ij}
\equiv m^2_i-m^2_j$, $\tilde{\Delta}^{}_{ij} \equiv
\tilde{m}^2_i-\tilde{m}^2_j$ and $\widehat{\Delta}^{}_{ij} \equiv
m^2_i-\tilde{m}^2_j$. In deriving Eq. (13), we have taken into
account the relationship\cite{Xreview}
\begin{eqnarray}
\sum_{i} \tilde{m}^2_i = \sum_{i} m^2_i +  \sum_{\alpha} {\cal
A}^{}_{\alpha \alpha} \ ,
\end{eqnarray}
which can be clearly seen from the traces of two sides of Eq. (4).
Note that only the mass-squared differences appear in our results,
and the terms in proportion to the absolute neutrino masses
disappear. This is consistent with the fact that oscillation
processes are irrelevant to the absolute neutrino masses. Eq. (13)
is very instructive and compact since it clearly shows the
connections between $|V^{}_{\alpha i}|$ and $|\tilde{V}^{}_{\alpha
i}|$. The survival probabilities denoted by $P(\nu^{}_{\alpha}
\rightarrow \nu^{}_{\alpha})$ or $P(\bar{\nu}^{}_{\alpha}
\rightarrow \bar{\nu}^{}_{\alpha})$\cite{Xreview} are only
sensitive to the moduli of $V(\tilde{V})$ and thus these
expressions will be very helpful for phenomenological analyses of
the disappearance processes in the neutrino oscillation
experiments.

For the case $\alpha \neq \beta$, we obtain the explicit
expressions
\begin{eqnarray}
\tilde{V}^{}_{\alpha i} \tilde{V}^{\ast}_{\beta i} =
\frac{1}{\prod \limits_{k \neq i} \tilde{\Delta}_{ik} } \left(
\sum_{j} F^{i j}_{\alpha \beta} V^{}_{\alpha j} V^{\ast}_{\beta j}
+ C^{}_{\alpha \beta} \right)  \ ,
\end{eqnarray}
where
\begin{eqnarray}
F^{ij}_{\alpha \beta} & = & \frac{1}{2}\left[ -
\widehat{\Delta}^{}_{ji} \left( {\cal A}^{2}_{\alpha \alpha} +
{\cal A}^{2}_{\beta \beta} + 4{\cal A}^{}_{\alpha \alpha}{\cal
A}^{}_{\beta \beta} \right) + \prod \limits_{k \ne i} \left( {\cal
A}^{}_{\alpha \alpha} + {\cal A}^{}_{\beta \beta} +
\widehat{\Delta}^{}_{j k}\right) + \prod \limits_{k \ne i}
\widehat{\Delta}^{}_{jk} \right] \ ,
\nonumber \\
C^{}_{\alpha \beta} & = & -\frac{1}{2} \sum_{\gamma,m,n}
\Delta^2_{mn} V^{}_{\alpha m}  V^{}_{\gamma n}  V^{\ast}_{\gamma
m} V^{\ast}_{\beta n} {\cal A}^{}_{\gamma \gamma} \ ,
\end{eqnarray}
and $\alpha \neq \beta$. In the limit ${\cal A} \rightarrow 0$,
our results are reduced to the vacuum case $\tilde{V} \rightarrow
V$.

To show the applications of our results, we define the rephasing
invariants in matter as
\begin{eqnarray}
\tilde{J}^{ij}_{\alpha \beta} \equiv {\rm Im}
\left(\tilde{V}^{}_{\alpha i}\tilde{V}^{}_{\beta
j}\tilde{V}^{\ast}_{\alpha j}\tilde{V}^{\ast}_{\beta i}\right) \ ,
\end{eqnarray}
which are similar to the one in the three-neutrino mixing
scheme\cite{Jarlskog}. Only nine of these invariants are
independent\cite{Guo}. We can choose any set of nine independent
$\tilde{J}^{ij}_{\alpha \beta}$ to fully describe CP and T
violation in matter.

The twelve unitarity quadrangles\cite{Guo}, which are the
geometrical presentations of the orthogonality of $V(\tilde{V})$
in the complex plane, are well related with CP and T violation in
neutrino oscillations. The twenty-four sides of six quadrangles
have been given in Eq. (16). Once precise measurements are
provided in future, we can systematically analyze the matter
effects on these quadrangle sides by taking use of our relations.

The probability of one neutrino $\nu^{}_{\alpha}$ converting to
another neutrino $\nu^{}_{\beta}$ in matter is given as
\begin{eqnarray}
P(\nu^{}_\alpha \rightarrow \nu^{}_\beta)  =  \delta_{\alpha\beta}
- 4 \sum_{i<j} \left [ {\rm Re} \left ( \tilde{V}^{}_{\alpha i}
\tilde{V}^{}_{\beta j} \tilde{V}^{\ast}_{\alpha j}
\tilde{V}^{\ast}_{\beta i} \right ) \sin^2 \tilde{F}_{ji} \right ]
- 2 \sum_{i<j} \left ( \tilde{J}^{ij}_{\alpha\beta} \sin 2
\tilde{F}_{ji} \right ) \ ,
\end{eqnarray}
where $\tilde{F}_{ji} \equiv 1.27 \tilde{\Delta}^{}_{ji} L/E$ with
$L$ standing for the baseline length (in unit of km) and $E$ being
the neutrino beam energy (in unit of GeV). From our simple and
exact formulae, the oscillation probabilities for all channels in
matter can be calculated and the CP-violating information can also
be extracted.

It should be mentioned that our formulae are also very suitable
for numerical analyses, and the relevant work will be elaborated
elsewhere.

\section{summary}

Motivated by the interesting extension of the standard flavor
mixing scheme of three active neutrinos with one sterile neutrino,
we have derived a full set of sum rules in the four-neutrino
mixing scheme. By using these sum rules, we have calculated the
effective neutrino mixing matrix in matter $\tilde{V}$ in terms of
$\tilde{m}^{}_i$, $m^{}_i$ and the genuine neutrino mixing matrix
$V$. The rephasing invariants, sides of six unitarity quadrangles
and oscillation probabilities in matter can be obtained by using
our formulae in Eqs. (13) and (16). These compact and simple
results are expected to be very helpful to describe the leptonic
flavor mixing and CP and T violation in matter in a
model-independent way. The relations are also very useful for a
systematic study of neutrino oscillations in the future
long-baseline experiments.

\vspace{0.5cm}

The author would like to thank Professor Z.Z. Xing for reading the
manuscript, making many corrections and giving a number of helpful
suggestions. He is also grateful to S. Zhou and W.L. Guo for
useful discussions. This work was supported in part by the
National Nature Science Foundation of China.

\end{document}